# Stochastic resonance in the growth of a tumor induced by correlated noises

ZHONG Weirong, SHAO Yuanzhi[1] & HE Zhenhui

Department of Physics, Sun Yat-sen University, Guangzhou 510275, China

Correspondence should be addressed to Shao Yuanzhi(E-mail: stssyz@zsu.edu.cn)

**Abstract** Multiplicative noise is found to divide the growth law of tumors into two parts in a logistic model, which is driven by additive and multiplicative noises simultaneously. The Fokker-Planck equation was also derived to explain the fact that the influence of the intensity of multiplicative noise on the growth of tumor cells has a stochastic resonance-like characteristic. An appropriate intensity of multiplicative noise is benefit to the growth of the tumor cells. The correlation between two sorts of noises weakens the stochastic resonance-like characteristic. Homologous noises promote the growth of the tumor cells.

  Since half a century ago, nonlinear physics has changed our views of world drastically [1]. People expected to use it to solve the complexity of biology [2, 3]. In the past decade, more and more evidence shows that noise plays an important role in nonlinear systems [4-10], such as noise-induced phase transitions, stochastic resonance (SR) and so on. Especially, SR uncovered a heap of complexities of biology [11, 12]. For many years, the law of tumors growth has been a challenging subject. Scientists try hard to find exact measures to control tumors and cure cancers with physical, mathematical, biological and chemical methods. And many models have been established, such as Eden model [13], Gompertzian growth model [14], self-limiting growth model [15], and feedback model [16]. Bru et al. [17] discovered that tumor growth is mainly governed by environmental pressures (host tissue pressure, immune response, etc.) through both simulations and experiments of the immune responses of mice. Logistic growth equation, a nonlinear equation with profound physical meaning, affords a good annotation about the origins and evolutions of biology [18]. Ai et al. [19] first applied logistic growth model induced by correlated noises to analyze the growth of the tumor cells and found that strong fluctuation in the growth rate would exterminate tumor cells. For the sake of simplicity, Ai et al. only dealt with some factors, including fluctuations of temperature, drugs, radiotherapy, etc., that have influence on the growth rate of the tumors. In fact, the environments affect the decay rate as well as the growth rate.

---

[1] Corresponding author: stssyz@zsu.edu.cn

In this paper, considering the fluctuation of the linear growth rate under a stable carrying capacity of environment, we study the influences of the intensities of correlated noises on the growth of the tumor cells, and indicate that multiplicative noise (MN) can divide the growth law of tumors into two parts. We will show that the intensity of multiplicative noise affects the growth of the tumor cells with an SR-like trend. The peak of SR depends upon the change of correlation between noises. We attempt to give an insight into the intrinsic growth principle of tumors and present a new concept for tumor treatments.

## 1 Growth model of tumors

Logistic model proposed by Verhulst [18] is written as

$$\frac{dx}{dt} = rx(1 - \frac{x}{K}) \qquad (1)$$

where $r$ is the linear per capita birth rate, $K$ is the carrying capacity of the environment and $x$ is the population. This is an ideal equation without fluctuation. Factually, tumor cells are always influenced by the environment, and their linear per capita birth rate changes and fluctuates consequentially. So it is reasonable to introduce noise $\xi(t)$ into the above model and rewrite $r$ as $r+\xi(t)$. Likewise, emigration and immigration of tumors, set at $\eta(t)$, also affect the change in the number of the tumor cells in a local area. The Langevin differential equation induced by noises is

$$\frac{dx}{dt} = rx(1-\frac{x}{K}) + x(1-\frac{x}{K})\xi(t) + \eta(t) \qquad (2)$$

Here $\xi(t)$ and $\eta(t)$ are respectively multiplicative and additive Gaussian white noises in the following forms:

$$\langle\xi(t)\rangle = 0, \qquad \langle\xi(t)\xi(t')\rangle = 2D\delta(t-t') \qquad (3a)$$

$$\langle\eta(t)\rangle = 0, \qquad \langle\eta(t)\eta(t')\rangle = 2Q\delta(t-t') \qquad (3b)$$

$$\langle\xi(t)\eta(t')\rangle = 2\lambda\sqrt{DQ}\delta(t-t') \qquad (3c)$$

in which $D$ and $Q$ are the intensities of the noises, and $\lambda$ denotes the strength of the correlation between $\xi(t)$ and $\eta(t)$, $\delta(t-t')$ is Dirac delta function at different moments. Since white noises in Eq.(2) are in accordance with Markov processes, the equivalent Fokker-Planck equation can be derived [6, 7]

$$\frac{\partial p(x,t)}{\partial t} = -\frac{\partial}{\partial x}[A(x)p(x,t)] + \frac{\partial^2}{\partial x^2}[B(x)p(x,t)] \qquad (4)$$

where $p(x,t)$ is the probability distribution function, $A(x)$ and $B(x)$ are

$$A(x) = rx(1-\frac{x}{K}) + Dx(1-\frac{x}{K})(1-\frac{2x}{K}) + \lambda\sqrt{DQ}(1-\frac{2x}{K}), \qquad (5a)$$

$$B(x) = Dx^2(1-\frac{x}{K})^2 + 2\lambda\sqrt{DQ}x(1-\frac{x}{K}) + Q, \qquad (5b)$$

Since the number of the tumor cells, $x$, is positive, according to the reflected boundary condition, the steady-state probability distribution (STPD) of Eq.(4) is given [20]

$$p_{st}(x) = \frac{N}{B(x)}\exp[\int^x \frac{A(x')}{B(x')}dx'] \qquad (6)$$

in which $N$ is a normalization constant.

The following parameters are defined to analyze the influence of noises on the tumors quantitatively.

$$\langle x\rangle = \int_0^\infty x p_{st}(x)dx, \qquad (7a)$$

$$\langle x^2\rangle = \int_0^\infty x^2 p_{st}(x)dx \qquad (7b)$$

$$\langle\delta x^2\rangle = \langle x^2\rangle - \langle x\rangle^2 \qquad (7c)$$

Here $<x>$ and $<\delta x^2>$ are the mean number of tumor cells and its variance,

respectively. The increase in $<x>$ means cells are growing, otherwise extincting. Augment of $<\delta x^2>$ denotes that the growth law of the cells is broken down, or is promoted contrarily.

## 2 Results and discussions

The growth and extension of a tumor are affected by the environment [17, 21]. The treatments have positive or negative effects on the environment of a tumor [22]. Therapies, such as chemotherapy and radiotherapy, are two-edged swords. They not only kill the tumors, but also activate them. Moreover, they can cause the canceration of the normal cells. All of these changes are undeterminate; therefore investigating the correlation of two sorts of noises, multiplicative and additive noises, is vital to the clarification of the underlying mechanism. In the model, the intensity of multiplicative noise refers to the strength of the treatment, i.e. the dosage of the medicine in chemotherapy or the intensity of the ray in radiotherapy. The intensity of additive noise serves as the capability of expansionary transfer of the tumors. The correlation between additive and multiplicative noises denotes the adaptability of the tumors to the treatments. If $\lambda=0$, the tumor is completely unconformable, or else if $\lambda=1$, the tumor develops a perfect adjustability.

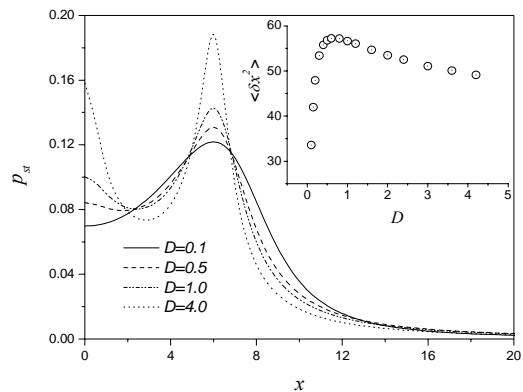

Fig.1 shows the STPD of the number of the tumor cells (NTC) under different intensities of MN. In Fig.1, the STPD peaks only at $x=K$ in the case of weak MN ($D=0.1$), indicating that the tumors grow steadily. However, the STPD changes from uni-peak to bi-peak states as the intensity of MN increases; one of the peaks appears at nearly $x=0$, and the other at $x=K$.

Fig.1. Dependence of the steady-state probability distribution on the intensity of multiplicative noise, $D$. Inset: Variance of the tumor cells number against $D$. The parameters are $\lambda=0.0$, $Q=2.0$.

The above change in curve trend suggests that there exists a phase transition in the growth law of the tumors driven by noises. The position of the peaks (PPs) near *x=0* denotes the tumor's deterioration, and PPs near *x=K* means the adaption of some tumor cells to the treatments. The influence of noises on the growth law of the tumors extends first and then shrinks with the increase of the intensity of MN , as shown in the inset of Fig.1. As a whole, an appropriate intensity of multiplicative noise can break down the growth law of the tumors, while superfluous noises improve their growth contrarily.

Fig.2 shows the change of NTC with the intensity of MN under different intensities of additive noises. As the intensity of MN increases, NTC increases first and then decreases, showing a typical SR-like characteristic. Obviously, an unsuitable intensity of the treatment can not kill tumors, but activate them. Consequently, in order to acquire a good cure of tumors, an adequately strong intensity of treatment is required. Unfortunately, the normal cells take a risk of canceration under excess intensity of treatment at the same time. This is also a dilemma for current chemotherapy and radiotherapy.

As the intensity of additive noise increases, the value of peak in Fig.2 rises but the position of peak remains intact. Fig.3 shows another SR-like characteristic under different correlations. The position of peak shifts with the changes in the correlations. The inset of Fig.3 displays NTC changes with the intensity of correlation at various intensities of noises.

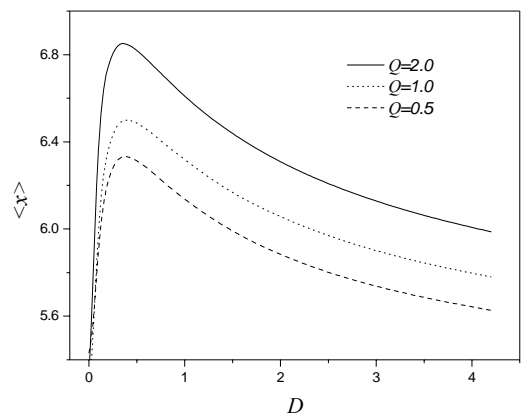

Fig.2. The varieties of the mean number of tumor cells with the intensity of multiplicative noise at $\lambda=0.0$, from bottom to top, $Q=0.5, 1.0, 2.0$,

In the case of weak noises, NTC drops with an increase in the correlation. If the noises are strong, NTC rises with an increase in the correlation. NTC does not change with the correlation at intermediate noise intensities.

These results suggest that a tumor grows and extends more steadily as the correlation becomes closer. This means that the tumor has a better adjustability to the treatment. When adjustability reaches a definite degree, i.e. $\lambda>0.8$, the peak of SR degenerates and disappears. Then the noise, however strong their intensities are, can not affect tumors, namely invalidation happens here. Consequently, a fixed treatment sometimes does nothing to remove tumors. Provided that the methods and intensities of the treatments are changed at intervals, better treatment effects are expected.

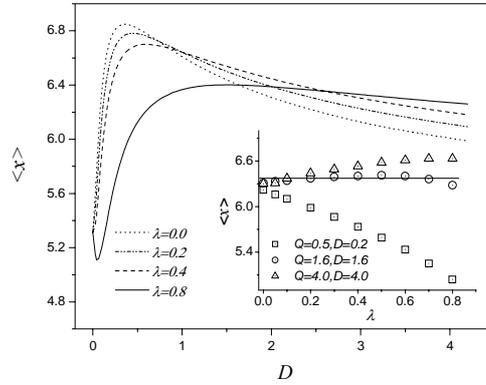

Fig.3. The varieties of the mean number of tumor cells with the intensity of multiplicative noise at different correlations. The parameter is $Q=2.0$. Inset: the relationship between $<x>$ and $\lambda$ under different intensities of noises.

## 3 Conclusions

There are two sorts of noises: one originates from environment like chemotherapy and radiotherapy, and the other roots in the tumors themselves. The former is multiplicative, and the latter is additive. The growth law of the tumors is divided into two parts under the effect of multiplicative noise. The influence of multiplicative noise on the tumors has a stochastic resonance-like characteristic. An appropriate intensity of multiplicative noise leads the tumor cells to develop steadily. The correlation between noises weakens the stochastic resonance-like characteristic. Homologous noises promote the growth law of the tumor cells.

## Acknowledgements

This work was partially supported by the National Natural Science Foundation of China (Grant No. 60471023) and the Natural Science Foundation of Guangdong Province (Grant No. 031554).

References

[1] Nicolis G. and Prigogine I., Self-organization in Nonequilibrium systems, New-York: Willey, 1977.

[2] Qi Anshen, Du Chanying, Nonliear model of immunity, Shanghai: Shanghai Scientific and Technological Education Publishing House,1998. 124~149 (in Chinese)

[3] Nicolis G. and Prigogine I., Exploring Complex, New-York: Freeman, 1986.

[4] Gammaitoni L., Hanggi P., Jung P., and Marchesoni F., Stochastic resonance. Rev. Mod. Phys., 1998, 70: 223~287

[5] Anishchenko V S, Astakhov V V, Neiman A B, Vadivasova T E, and Schimansky-Geier L., Nonlinear Dynamics of Chaotic and Stochastic Systems. Berlin Heidelberg: Springer-Verlag, 2002. 327~363

[6] Gardiner C. W., Handbook of Stochastic Methods for Physics, Chemistry and the Natural Science, Berlin: Springer-Verlag, 1983.

[7] Hu G., Stochastic Forces and Nonlinear Systems, Shanghai: Shanghai Scientific and Technological Education Publishing House, Shanghai, P. R. China, 1994. (in Chinese)

[8] Zaikin A A, Kurths J, and Schimansky-Geier L. Doubly Stochastic Resonance. Phys. Rev. Lett., 2000, 85: 227~231

[9] Jia Y. and Li J. R., Steady-state analysis of a bistable system with additive and multiplicative noises. Phys. Rev. E, 1996, 53: 5786~5792

[10] Hanggi P and Riseborough P. Activation rates in bistable systems in the presence of correlated noise. Phys. Rev. A, 1983, 27: 3379~3382

[11] Dean Astumian R., Adair R. K., and Weaver James C., Stochastic resonance at the single-cell level, Nature, 1997, 388: 632~633

[12] Russell D. F., Wilkens L. A., and Moss F. Use of behavioural stochastic resonance by paddle fish for feeding. Nature, 1999, 402: 291~294

[13] Yu J, Hu G., and Ma B K. New growth model: The screened Eden model. Phys. Rev. B, 1989, 39: 4572~4576

[14] Molski M and Konarski J. Coherent states of Gompertzian growth. Phys. Rev. E, 2003, 68: 021916-1~7

[15] Kar S., Banik S. K., and Ray D. S. Class of self-limiting growth models in the presence of nonlinear diffusion. Phys. Rev. E, 2002, 65: 061909-1~5

[16] Scalerandi M. and Sansone B. C. Inhibition of Vascularization in Tumor Growth. Phys. Rev. Lett., 2002, 89: 218101-1~4

[17] Bru A., Albertos S., Garcia-Asenjo J. A. L., and Bru I. Pinning of Tumoral Growth by Enhancement of the

Immune Response. Phys. Rev. Lett., 2004, 92: 238101-1~4

[18] Verhulst P. -F.. Recherche mathèmathiques sur le loi d'accroissement de la population. Nouveau Memoirės de l'Acadèmie Royale des Sciences et Belles Lettres de Bruxelles, 1845, 18: 3~38

[19] Ai B. Q., Wang X. J., Liu G. T., and Liu L. G.. Correlated noise in a logistic growth model, Phys. Rev. E, 2003, 67: 022903-1~3

[20] Wu D. J., Li C., and Zhi K. S. Bistable kinetic model driven by correlated noises: Steady-state analysis. Phys. Rev. E, 1994, 50: 2496~2502

[21] Bru A., Albertos S., Subiza J. L., Garcia-Asenjo J. L., and Bru I. The Universal Dynamics of Tumor Growth. Biopys. J., 2003, 85: 2948~2961

[22] Ferreira S. C. Jr, Martins M. L., and Vilela M. J. Morphology transitions induced by chemotherapy in carcinomas in situ. Phys. Rev. E, 2003, 67: 051914-1~9